\pdfoutput=1
\documentclass[aps,prl,preprint,superscriptaddress]{revtex4-1}
\usepackage{dcolumn}
\usepackage{bm}
\usepackage{amsmath,amssymb,graphicx,color}
\usepackage{hyperref}
\usepackage[english]{babel}

\hypersetup{bookmarksnumbered=true,	unicode=false,	pdfstartview={FitH}, pdfnewwindow=true, colorlinks=true,linkcolor=blue, citecolor=blue, filecolor=blue, urlcolor=blue}

\begin{document}

\title{How Material Heterogeneity Creates Rough Fractures}

\author{Will Steinhardt}
\affiliation{Department of Earth and Planetary Sciences, Harvard University, Cambridge, MA,	02138, USA}
\author{Shmuel M. Rubinstein}
\affiliation{The Racah Institute of Physics, Hebrew University, Jerusalem, Israel}

\begin{abstract}

Fractures are a critical process in how materials wear, weaken, and fail whose unpredictable behavior can have dire consequences.  While the behavior of smooth cracks in ideal materials is well understood, it is assumed that for real, heterogeneous systems, fracture propagation is complex, generating rough fracture surfaces that are highly sensitive to specific details of the medium.  Here we show how fracture roughness and material heterogeneity are inextricably connected via a simple framework. Studying hydraulic fractures in brittle hydrogels that have been supplemented with microbeads or glycerol to create controlled material heterogeneity, we show that the morphology of the crack surface depends solely on one parameter: the probability to perturb the front above a critical size to produce a step-like instability. This probability scales linearly with the number density, and as heterogeneity size to the $5/2$ power. The ensuing behavior is universal and is captured by the 1D ballistic propagation and annihilation of steps along the singular fracture front.

\end{abstract}

\maketitle{}

As a crack moves through a material, it leaves in its wake a fracture surface that preserves the time history of the shape of the crack front.  For many materials this surface is rough, a remnant of the crack’s tortuous path. Roughness is intrinsically three-dimensional, indicating the distortion of a spatially-extended, singular, crack tip line.  Idealized fractures advance when the flux of released elastic stress at the crack tip is greater than the energetic cost of breaking the material and generating new surface \cite{lawn1993}, therefore, rougher cracks require more energy to produce. Once cracks exists within a material, they fundamentally alter its characteristics, with their roughness affecting properties including fluid transport \cite{tsang1983,aydin2000,bunger2017} and frictional stability \cite{brodsky2011,brodsky2007,candela2011,dieterich1994,greenwood1966,persson2001}.

For slow brittle fractures, the crack front is largely straight and forms flat, mirror-like surfaces.  Nevertheless, a close inspection of these surfaces often reveals the presence one of the most prominent components of brittle fracture roughness: long, step-like discontinuities called step lines.  This class of fractographic features is found in a wide array of both hard \cite{sommer1969,hull1995,cooke1996} and soft \cite{baumberger2008,tanaka1996,tanaka1998,kolvin2017,kolvin2018,goldstein2012} brittle materials.  Unlike microbranches, which appear at significant fractions of the Rayleigh wave speed \cite{sharon1996}, step lines form on slow, quasi-static fractures as well.

Step lines arise from a critical twisting, or mixed mode I + III loading, of the fracture front \cite{sommer1969}.   Rather than reorient the entire front, an instability mediates the formation of a step-like defect, locally accounting for the twist, while the rest of the crack remains flat \cite{sommer1969,ronsin2014,lazarus2001,lazarus2001a,leblond2019,chen2015,pons2010,lin2010,pham2016,pham2017,kolvin2018}.  Mixed mode loading at the crack tip can result from applied boundary conditions \cite{sommer1969,lazarus2001,lazarus2001a,hull1993,hull1995,hull1999}, but is also understood to arise from material heterogeneity \cite{leblond2016,ravi-chandar1984}.  However, the connection between heterogeneity and roughness has not been resolved because it is difficult to systematically control or measure heterogeneity in an experimental sample.  Moreover, it is assumed that rough fracture surfaces result from complex crack behavior that is highly sensitive to the details of the medium.

In this work we show that heterogeneity and fracture roughness are inextricably linked through a simple framework where the evolution of the system depends solely on one parameter: the probability that the heterogeneity perturbs the front enough to produce a step line. By analyzing the broken surfaces of hydrogels prepared with controlled, discrete heterogeneities, we show that increasing either the size or amount of heterogeneities present leads to a marked increase in the density of step lines on the fracture surface.  After step lines form, they continuously drift laterally along the front \cite{tanaka1998,kolvin2018}, leading them to interact in a manner that, on average, reduces the number of steps.  Eventually, step creation and annihilation rates balance, leading to a steady state. We identify the relationship between the scale of a heterogeneity and the probability of step nucleation on a crack front, connecting the material heterogeneity to the excess energy required to propagate a crack.  We also show that our framework can be extended to complex backgrounds of non-discrete, inhomogeneity, indicating that the resultant roughness of a fracture is invariant to the specific process that perturbs the crack front. Finally, when heterogeneities are present at high densities and on many different scales, interactions between steps and the complex background result in fracture surfaces resembling those observed in natural systems.

\begin{figure}
 \includegraphics[scale=0.3]{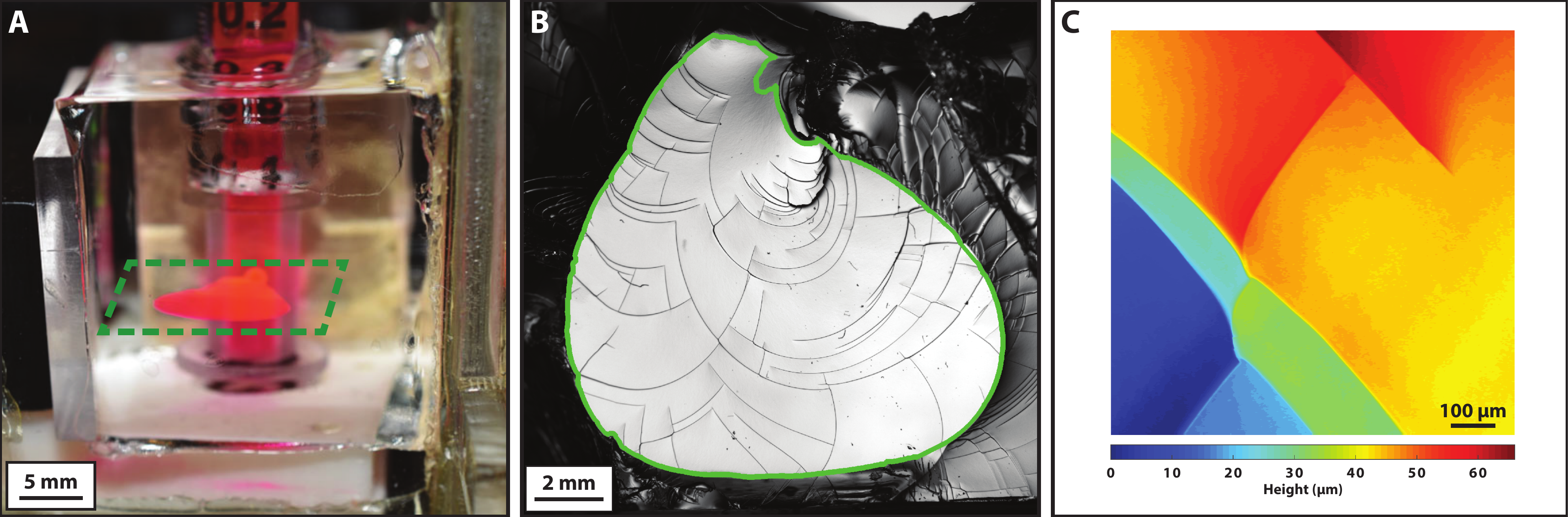}
	\caption{(A) Image of typical fracture in a hydrogel after crack propagation has stopped.  The resultant penny-shaped crack (green square) in a hydrogel within our experimental apparatus.  The fracture is broadly flat and smooth.  (B)  Microscopy image of typical fracture surface segmented by many arched step lines.  There is a clear boundary (green line) between the surface created by controlled hydraulic fracture (inside) and surface created when separating the gel for imaging (outside) (C) Typical example of surface height map measured with confocal microscope showing that steps generate additional surface area and create topographic boundaries.
	}
	\label{setup}
\end{figure}

Hydraulic fractures in hydrogels are an excellent model system in which to study fracture mechanics. Hydraulically-driven fractures generate large, reproducible surfaces \cite{sommer1969}, while hydrogels offer unique flexibility in tuning material properties and geometry.  In this study, experiments are performed by flowing a dyed fluid into a small pre-fracture in a 1x1x1 inch brittle, chemically cross-linked hydrogel (for details, see Section S3).  This fracture then grows until it reaches the sample edge, as shown in Figure \ref{setup}A.

While the gel fracture surfaces are nominally flat and smooth, they are pervasively segmented by long curved step lines, which separate two otherwise largely flat planes, as shown in Figure \ref{setup}B and C.  When steps nucleate they drift laterally along the front, either left or right, at a rate equal to that of crack propagation, causing step lines to maintain a 45-degree angle to the crack front \cite{tanaka1998,kolvin2018}.  For the curved crack fronts in our experiment, this results in arched step lines.  Steps drift in both directions along the front, leading to pairwise interactions as they meet.  These interactions result in zero, one, or two outgoing steps, and can only reduce the number of steps. Thus, nucleations increase the number of steps on the front, while interactions, on average, reduce the number of steps.

\begin{figure}
 \includegraphics[scale=0.55]{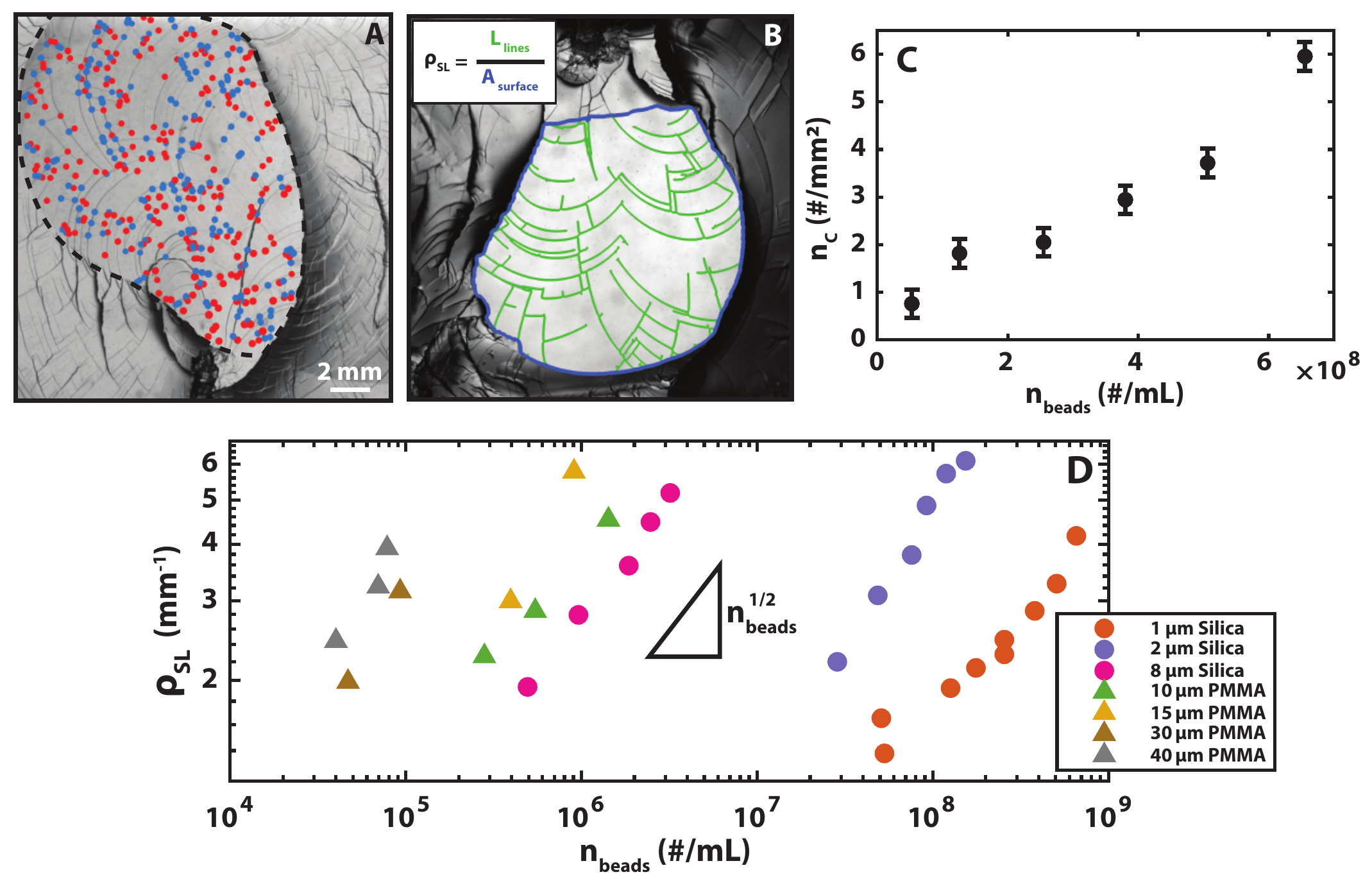}
	\caption{
	(A) Typical fracture surface with nucleations (red dots) and interactions (blue dots) indicated within experimental fracture area (black dashed line). (B) Typical analyzed fracture surface. Image processing is used to create a skeletonized map of step lines (green) for calculating $\rho_{\mathrm{SL}}$.  (C) Step creations per unit area, $n_{c}$, for 1 $\mu$m silica beads as a function of volumetric bead density, $n_{\mathrm{beads}}$.  Error bars represent uncertainty in measurement of nucleations (D) $\rho_{\mathrm{SL}}$ as a function of $n_{\mathrm{beads}}$ for bead sizes ranging from 1-40 $\mu$m.  Each appears to grow as roughly as $n_{\mathrm{beads}}^{1/2}$. Note that the number of beads required to generate an equivalent $\rho_{\mathrm{SL}}$ decreases drastically with increasing bead size.
	}
	\label{plots}
\end{figure}

For each surface we measure the number of step nucleations, (including new steps that branch off existing steps) per unit area, $n_{C}$, as well as the total length of lines, or mileage, per unit area on the fracture surface (excluding the region near the crack origin), $\rho_{\mathrm{SL}}$, as shown in Figure \ref{plots}A and B.  Step lines reach a steady-state height \cite{kolvin2018}, thus $\rho_{\mathrm{SL}}$ is equivalent to the excess surface energy required to generate the fracture compared to a perfectly smooth crack.

When the gel is supplemented with a low number ($<$1 wt$\%$) of rigid microbeads, acting as local heterogeneities, the density of step lines on the surface increases significantly.  Systematically increasing the number of beads, $n_{\mathrm{beads}}$, leads to a monotonic increase in both the number of step nucleations, and $\rho_{\mathrm{SL}}$, by up to a factor of six, as shown in Figure \ref{plots}C and D.

The number of discrete steps created per unit area, $n_{c}$, grows linearly with $n_{\mathrm{beads}}$, as shown in Figure \ref{plots}C, indicating that the probability of a step nucleating along the front, $P$, scales linearly with the density of heterogeneities.  By comparison, $\rho_{\mathrm{SL}}$ grows sub-linearly for a range of bead sizes between 1-40 $\mu$m and does not depend on the material properties of the beads. The number of beads required to generate a similar $\rho_{\mathrm{SL}}$ decreases significantly with increasing radius, $r$, as shown in Figure \ref{plots}D.  This indicates that $P$ is a strong function of $r$, and thus, takes the form

\begin{equation}
    P \propto n_{\mathrm{beads}} f(r) \tag{1}\label{eq:1}
\end{equation}

Step interactions require two steps and, on average, result in fewer than two outgoing steps. By contrast, when the front interacts with a bead, it nucleates a step, increasing the number of steps on the front.  $P$ is proportional to the density of beads, suggesting that the beads are well dispersed, and thus act independently. By contrast, the probability of an interaction between steps increases with the density of steps present on the front at a given time, $\rho_{\mathrm{steps}}$.  

\begin{figure}
 \includegraphics[scale=0.5]{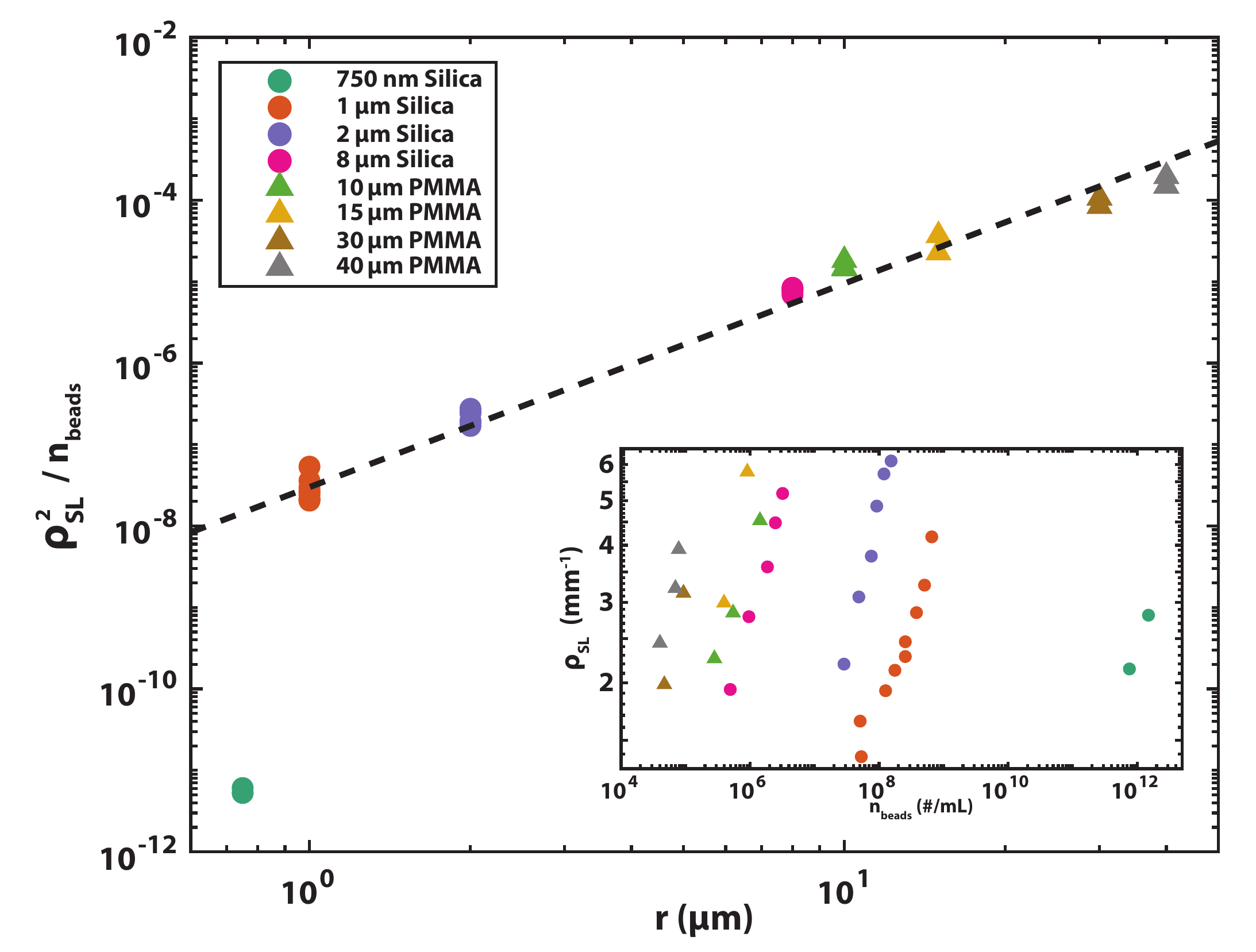}
	\caption{$\rho_{\mathrm{SL}} / n_{beads}^2$ as a function of bead radius, $r$.  Rescaling collapses all data between 1-40 $\mu$m onto a single line.  The dashed line indicates a slope of 5/2. The scaling regime breaks down sharply for r $<$ 1 $\mu$m. (insert) $\rho_{\mathrm{SL}}$ as a function of $n_{beads}$, similar to Figure 2E, that includes the r = 750 nm data.
    }
	\label{collapse}
\end{figure}

Steps are confined to move continuously on the singular fracture front. Thus, from a statistical physics point of view, our system is equivalent to ballistic particles in one dimension. If the locations and drift directions of steps are random, the probability of two steps interacting at a given time should scale as $\rho_{\mathrm{steps}}^2$.  This leads to a framework where as a crack front propagates, the creation rate remains constant while the probability of interactions grows.  Because interactions reduce the number of steps, the interaction rate will continue to grow until the rate of nucleations and annihilations through interaction are equal.  As a result, the system reaches a steady-state density of steps, $\rho_{\mathrm{ss}}$, such that $\rho_{\mathrm{ss}}^2$ is proportional to $P$.  Indeed, outside of the immediate area where the fracture initiates, the density of steps lines is broadly consistent, as shown in Figure S1.  Combining this with Eq. \ref{eq:1}, we find:

\begin{equation}
  \rho_{\mathrm{ss}}^2 \propto n_{\mathrm{beads}} f(r) \tag{2}\label{eq:2}
  \end{equation}

When the front advances a fixed distance, $L$, each step creates the same amount of mileage, $L\sqrt{2}$, due to the constant front angle. Thus, the rate at which mileage is generated is proportional to the instantaneous number of steps on the front.  Assuming the total mileage is measured on a surface produced predominantly at steady state, $\rho_{\mathrm{SL}} \propto \rho_{\mathrm{ss}}$.  Thus,

\begin{equation}
  {\rho_{\mathrm{SL}}^2  \propto n_{\mathrm{beads}}  f(r)} \tag{3}\label{eq:3}
\end{equation}

or 

\begin{equation}
  {\frac{ \rho_{\mathrm{SL}}^2} {n_{\mathrm{beads}} }  \propto  f(r)}  \tag{4}\label{eq:4}
\end{equation}

Plotting $\rho_{\mathrm{SL}}^2/n_{\mathrm{beads}}$  as a function of $r$ collapses the data onto a single line, as shown by Figure \ref{collapse}.  This suggests that by neglecting the area around crack initiation, we capture a surface that is predominately at steady state.  In addition, the slope of this line indicates that $f(r)$ scales as $r^{5/2}$.  Therefore the complete scaling function for the probability of nucleating a step is given by:

\begin{equation}
  P  \propto n_{\mathrm{beads}} r^{5/2}  \tag{5}\label{eq:5}
\end{equation}

And for $\rho_{\mathrm{SL}}$, a measure of the excess surface energy due to heterogeneity, is given by:

\begin{equation}
  \rho_{\mathrm{SL}}  \propto \sqrt{n_{\mathrm{beads}}} r^{5/4}  \tag{6}\label{eq:6}
\end{equation}

The scaling regime for $P$, described by Eq \ref{eq:6}, breaks down sharply below $r = 1 \mu$m, with the rescaled data for 750 nm beads falling 4 orders of magnitude below the trend for 1-40 $\mu$m beads.  Previous work has shown that a critical ratio of mode I/III loading is required to generate a step line \cite{sommer1969}.  We therefore suspect that beads below ~1 $\mu$m do not generate a large enough perturbation of the front to nucleate a step.  Instead, the minor roughness increase we do observe is likely the result of bead clustering.  Surprisingly, this ~1 $\mu$m cutoff indicates the presence of a length scale for a system governed by linear elastic fracture mechanics, which has classically considered to be scale-invariant.  This is consistent with recent work showing that heterogneities significantly smaller than the size of the fracture process zone do not affect the dynamics or fracture energy of a crack \cite{taureg2020}.

This framework fully captures the connection between heterogeneity and surface roughness for systems with well defined, discrete heterogeneities. However, a key assumption for quantifying the size and number of heterogeneities is that they are well distributed, and as a result, the front is locally straight.  Many natural systems do not have localized, dominant defects, and instead have smoother backgrounds of inhomogeneity, exhibiting non-local correlations on multiple scales. In these systems step nucleation and interactions may not necessarily be a local process.  Our framework relates the heterogeneity of the intact material to the excess energy required to generate a rough surface through the probability of nucleating steps, and thus does not necessitate specific knowledge of the nature of the heterogeneity. We show this by analyzing hydrogels with imhomogeneity that is controlled chemically.

Gels without any beads still produce a small number of step lines.  The value of $\rho_{\mathrm{SL}}$ for these experiments, $\rho_{\mathrm{SL}}^0$ , is approximately $1.13 \pm 0.1 mm^{-1}$.  There are a number of proposed mechanisms that cause structural heterogeneity in hydrogels \cite{seiffert2017}, including differences in the rate of monomer and crosslinker consumption during gelation, local fluctuations in polymer concentration, which become "frozen" in the polymer network during gelation\cite{sakai2014}, and microsyneresis. We theorized that the addition of a second phase to the solvent could accentuate these effects, creating inhomogeneity in polymer density that is more gradual and does not exhibit a sharp phase boundary compared to the highly-localized rigid beads. 

\begin{figure}
 \includegraphics[scale=0.6]{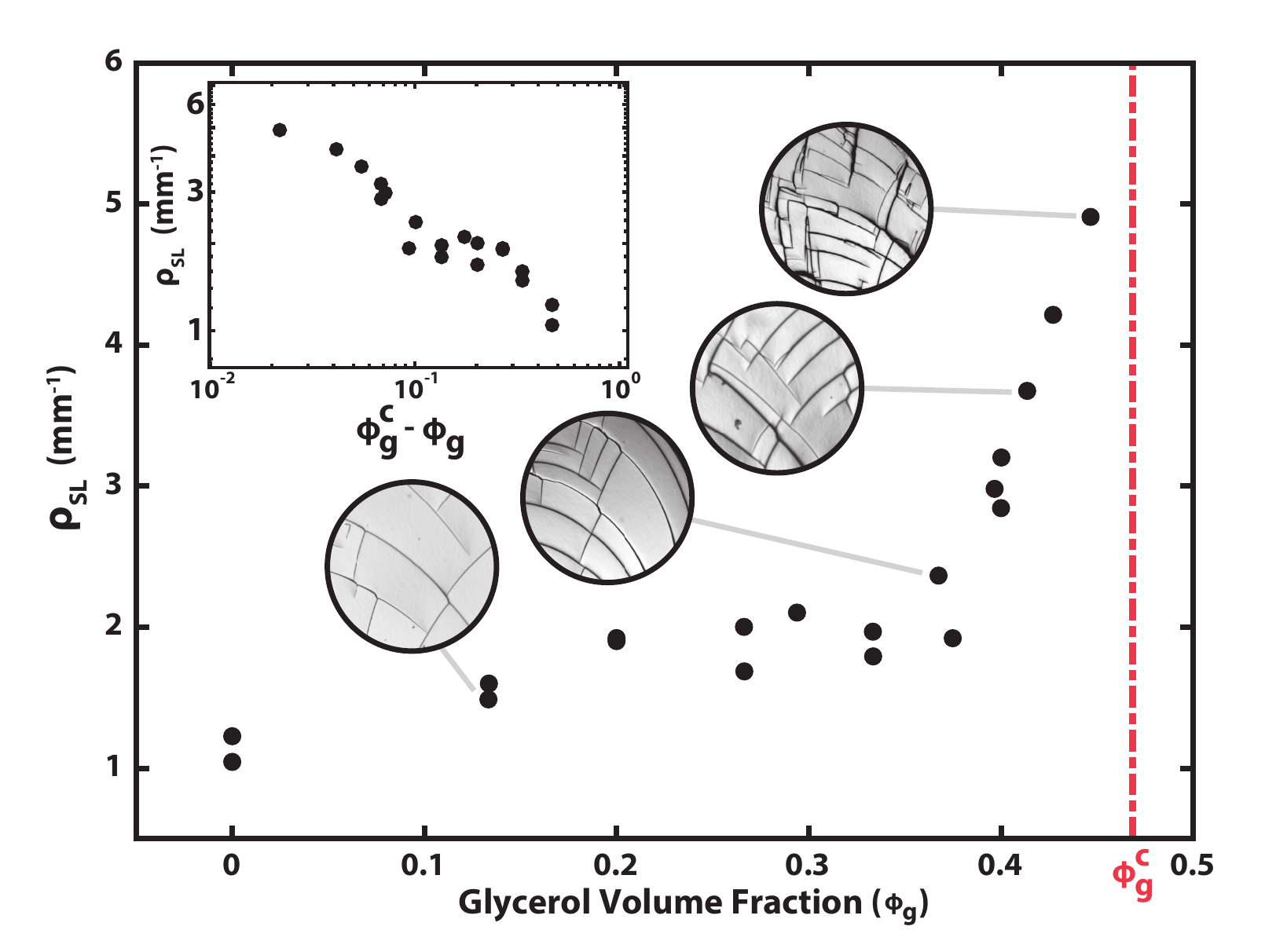}
	\caption{ $\rho_{\mathrm{SL}}$ as a function of glycerol volume fraction in hydrogel solvent, $\phi_{g}$. Typical images of fracture surfaces for corresponding samples are shown.  Each circle is 2.5 mm in diameter. Dashed, red line, indicates the solubility limit of PEGMEMA in glycerol-water solutions, $\phi_{g}^{c}$.  (Inset) $\rho_{\mathrm{SL}}$ as a function of distance from the solubility limit.
	}
	\label{glycerol}
\end{figure}

Indeed, when gels are polymerized in a solvent of glycerol and water, there is a continuous increase in $\rho_{\mathrm{SL}}$ up to a factor of 5 as a function of distance from the critical solubility limit, $\phi_{g}^{c}$, (46.5\% for PEGMEMA in glycerol-water), as shown in Figure \ref{glycerol}.  This is a clear indication that solubility drives the formation of the inhomogeneity. However, even very close to the solubility limit, the gels are fully transparent, showing no optical aberrations indicative of this inhomogeneity, and attempts to measure either the scale or amount of heterogeneity present via static light scattering were unsuccessful, making a full characterization impossible \footnote{Attempts to measure heterogeneity of glycerol-supplemented hydrogels via static light scattering were unsuccessful due to natural heterogeneity present in the base materials. \textit{Private communications with S. Seiffert}}.  
 
\begin{figure}
 \includegraphics[scale=0.6]{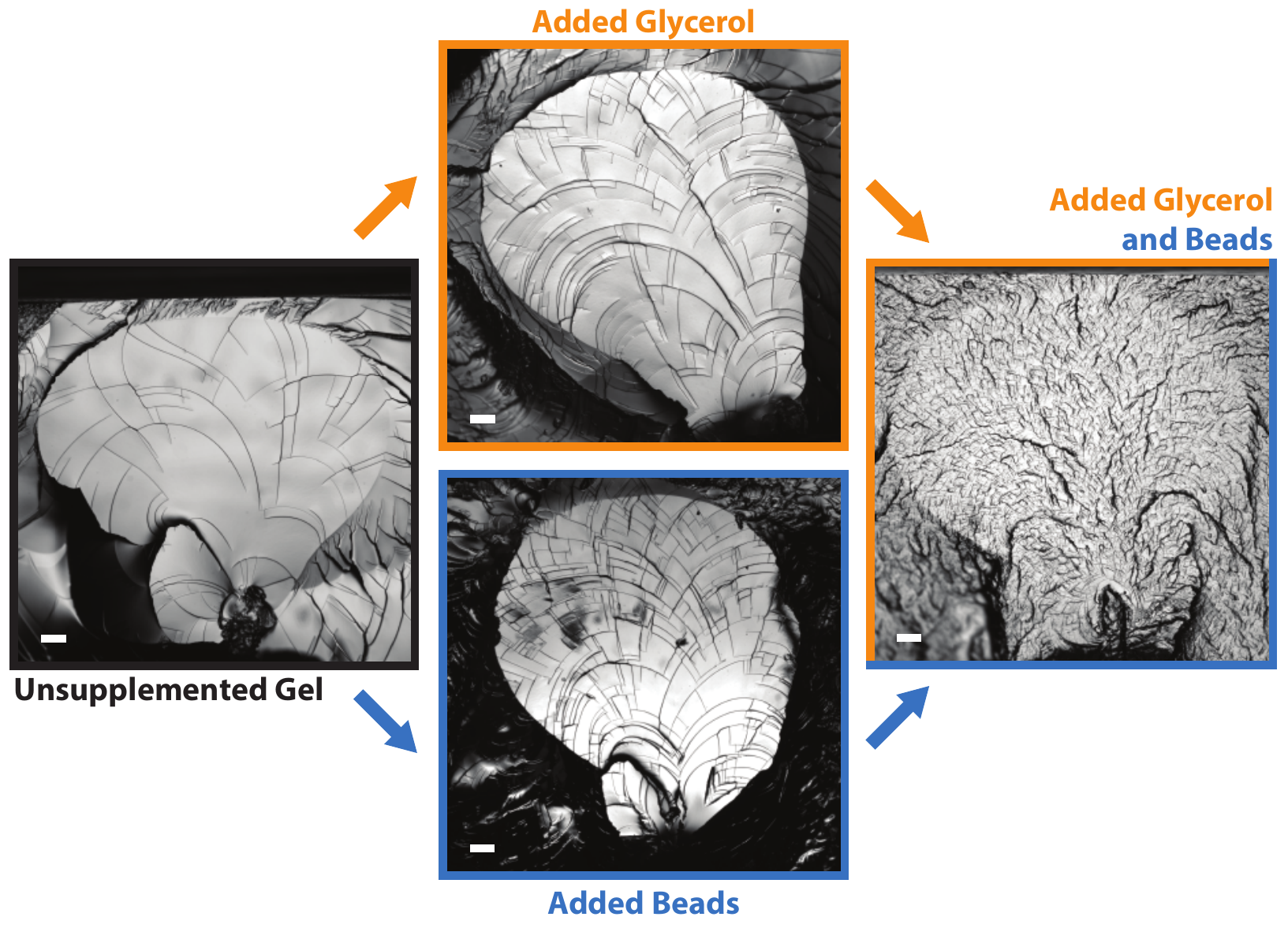}
	\caption{
	Typical images of fracture surfaces in hydrogels without any added heterogeneity (black outline), with added microbeads which act as localized heterogeneities (blue outline), and by adding glycerol to the solvent before polymerization (orange outline), which creates a more gradual background of inhomogeneity. When both glycerol and beads are used, creating complexity of multiple types and scales, significantly rougher surfaces (right) are generated that resemble those seen in natural materials.  Fracture propagation is approximately from the bottom to the top of each image.  White scale bars in bottom left are 1 mm.
	}
	\label{2paths}
\end{figure}

Surprisingly, the fracture surfaces of glycerol-supplemented gels are indistinguishable from those in bead-supplemented ones, as shown in Figure \ref{2paths}, suggesting that fracture roughness does not depend on the nature of the heterogneity or the specific process that perturbs the front, only on the probability of nucleating a step.

Steps produced in unsupplemented gels are caused by a low amplitude background of inhomogeneity.  A large enough bead creates a strong localized perturbation to the stress field, which overwhelms the subtle effect of the background. Indeed, the scaling discussed above for the bead-supplemented hydrogels does not include $\rho_{\mathrm{SL}}^0$, suggesting that in this regime, the stability of the fracture front is set exclusively by the dominant perturbation. However, as the amplitude of the background increases, a new regime emerges where neither discrete heterogeneity created by beads, nor more gradual inhomogeneity due to glycerol dominate.  The resulting fracture surfaces look fundamentally different, and resemble those observed in more complex brittle materials \cite{bahat1991,christensen2019,kimura1977,issa1994,carpinteri1999}, as shown in Figure \ref{2paths}.  This complex fracture morphology may be due to beads and/or steps interacting through the background, or  additionally, at high enough densities of steps along the front, interactions may no longer be simple and pairwise.  Regardless, realistic fracture topographies can be generated through the creation and interaction of step lines, potentially leading to previously observed self-affine fracture roughness \cite{mandelbrot1984,ponson2006}.

We have developed a simple model that connects material heterogeneity to fracture roughness.  For ideal, brittle materials, linear elastic fracture mechanics successfully predicts crack propagation, but it was commonly assumed that for heterogeneous materials, fractures are extremely sensitive to the details of the medium.  We have shown that at least for quasi-static fractures, this is not the case.  Instead, the evolution of the system depends solely on the probability to perturb the front enough to produce a step. The ensuing behavior is universal and results from the effectively 1D ballistic propagation and annihilation of steps along the fracture front.

The evolution of step lines on crack fronts is nearly identical to the seemingly unrelated system of decay kinetics in chemically reactive ideal gases. Extensively examined since the 1980s and named \textit{ballistic annihilation}, this theoretical framework considers the annihilation reaction $A+A \xrightarrow{} 0$ of ballistic particles. In one dimension, an exact solution \cite{elskens1985} shows that the concentration of ballistic particles decays as $1/\sqrt{t}$, while a mean field approximation predicts a $1/t$ scaling.  This discrepancy highlights that this seemingly simple system is in effect quite difficult to analyze \cite{biswas2021,burdinski2019}, hinting at why an understanding of fracture roughness has been so elusive. While the theoretical analysis of this system has been extended to partial annihilation \cite{kafri2001}, to our knowledge, there has been no investigation of a system that includes the creation of ballistic particles. As a result, the possibility of reaching a steady state has not been considered.  

Linear elastic fracture mechanics is a scaleless theory, yet prior observations have provided strong indirect evidence for the existence of a material-dependent length scale \cite{dugdale1960,chen2017,taureg2020,livne2008,kolvin2017}.  It has been suggested that this length is related to the size of the fracture process zone and is important in determining the toughness of soft materials\cite{chen2017}. The sharp breakdown in scaling for beads below 1 $\mu$m in our system is the most direct experimental evidence yet for a critical microscopic length scale in brittle fracture mechanics.

Our framework demonstrates how steady state fracture roughness results from a balance between step nucleations and interactions, providing a scaffold for a complete theory that would require detailed descriptions of each behavior. However, these processes are inherently three-dimensional and dynamic, necessitating both measurement and theory that capture this dimensionality.  Importantly, it is impossible to reduce the dimensionality of the system to a single 2D plane, highlighting the distinction between mixed mode I/II \cite{gao1989,gao1991,rice1994,taureg2020} and the current case of mixed mode I/III loading \cite{pons2010,leblond2019,lazarus2001,lazarus2001a}.  The complexity of step interactions arises from the fact that in spite of their intricate three-dimensional morphology\cite{sommer1969,hull1995,baumberger2008,kolvin2018a}, steps are topologically bound to the same one-dimensional fracture front-line, constraining their motion and interactions with other steps \cite{tanaka1998}.  In addition, these interactions occur dynamically during crack growth, so their geometry may also affect both local and global fracture propagation. A full description of the rules and dynamics that govern step nucleation and interaction, completing the framework presented above, will offer a pathway to fully understanding of how real materials break.

\section{Acknowledgements}

The authors would like to thank Tom Kodger, Dmitry Garagash,  and Robert Viesca for helpful discussions, and Sebastian Seiffert for preliminary optical measurements of hydrogel heterogeneity. This work was supported by the National Science Foundation through the Harvard Materials Research Science and Engineering Center (DMR-1420570).  S. M. R. acknowledges support from the Alfred P. Sloan Research Foundation (FG-2016-6925).




%

\end{document}


\title{Supplementary Materials for How Material Heterogeneity Creates Rough Fractures}

\author{Will Steinhardt}
\affiliation{Department of Earth and Planetary Sciences, Harvard University, Cambridge, MA,	02138, USA}
\author{Shmuel M. Rubinstein}
\affiliation{The Racah Institute of Physics, Hebrew University, Jerusalem, Israel}

\maketitle{}

\section{Distribution of Step Line Density}

\begin{figure} [htp]
 \includegraphics[scale=0.6]{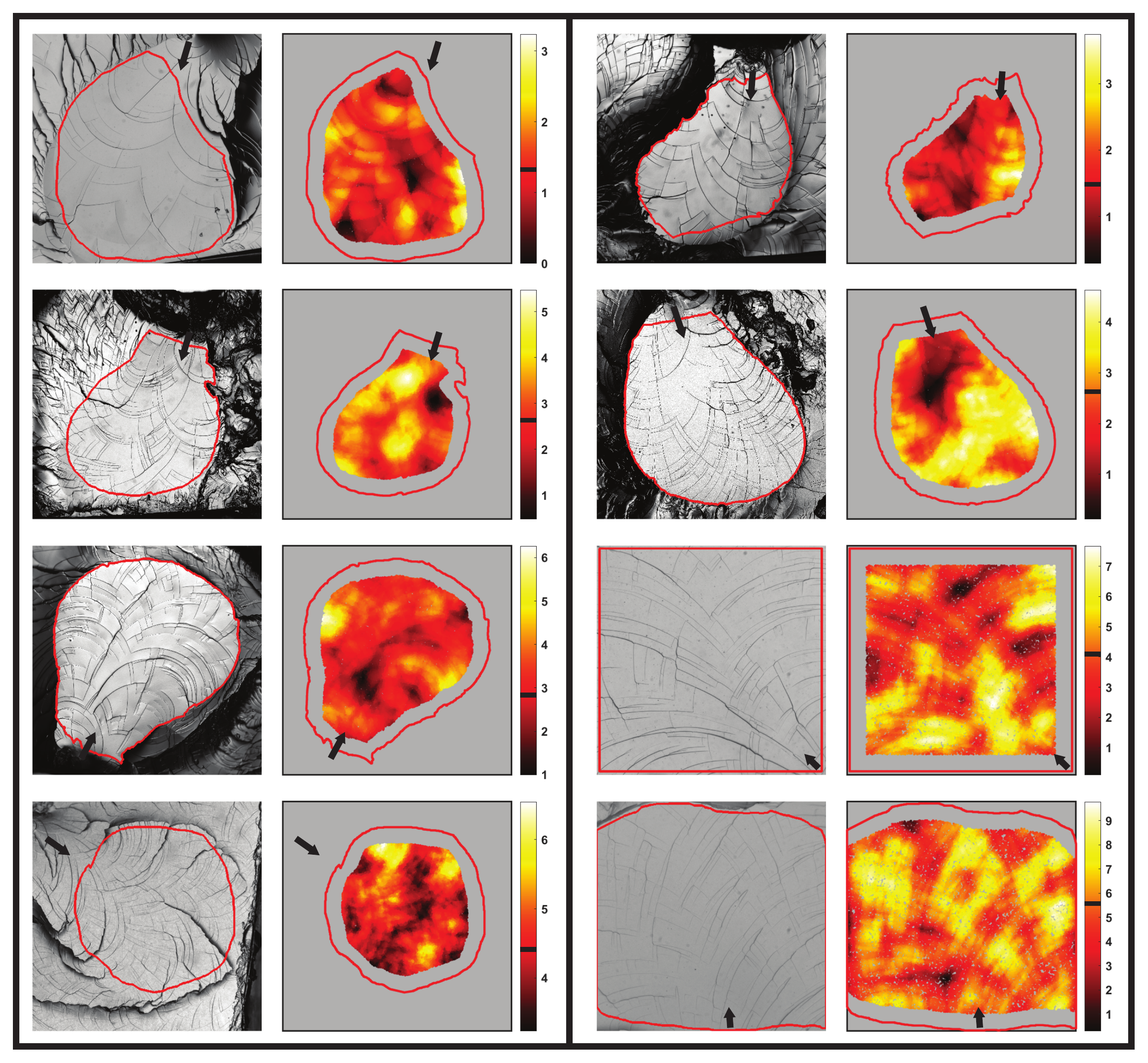}
	\caption
	{
	    \textbf{Fracture surfaces achieve a steady state. } Eight images (left columns) of fracture surfaces with corresponding heatmaps (right columns) of $\rho_\mathrm{SL}$ within the masked fracture surface (red line).  Heatmaps are generated by selecting 30000 random locations on the surface that are $>$150 pixels away from the edge of the mask and measuring $\rho_\mathrm{SL}$ within a circle with $d = 300$ pixels.  Black arrows indicate the approximate direction of crack propagation away from the fracture origin.  While $\rho_\mathrm{SL}$ varies over the surface, these variations are broadly consistent and do not increase or decrease monotonically with distance from the initiation point, indicative of a surface generated by a front at steady state.  Note that fractures are propagated at a constant flow rate and thus slow with increasing radius.  However, there is no radial dependence in $\rho_\mathrm{SL}$, indicating that step generation is not rate dependent for the slow, quasi-static fractures in this experiment.
	}
	\label{heatmaps}
\end{figure}

\section{Details of Surface Line Measurement}

Images are taken on a Zeiss Axiozoom V16 in brightfield mode, and are lit to enhance contrast between the steps and surrounding gel.  Images have a spatial resolution of ~8 microns/pixel.  This is more than sufficient to clearly identify each individual line, and the resulting analysis is not sensitive to the details of how it is performed.  In the standard analysis procedure, the images are thresholded, binarized, and skeletonized to create maps of step lines.  Additional morphological operations are performed including erosion, dilation, and spurring, followed by a visual inspection to ensure the accuracy of the mapping with the map being altered by hand to remove erroneous lines and add missed lines.  When this process fails to capture the lines well, lines are fully mapped by hand.